\documentclass[fleqn,12pt,twoside]{article}
\usepackage[headings]{espcrc1}
\readRCS
$Id: espcrc1.tex,v 1.2 2004/02/24 11:22:11 spepping Exp $
\usepackage{times}
\usepackage{graphicx}
\usepackage{footnote}

\title{\bf Particle production at very low and intermediate transverse momenta in d+Au and Au+Au collisions}

\author{ Adam Trzupek$^3$ for the PHOBOS Collaboration\\[2mm]
B.Alver$^4$,
B.B.Back$^1$,
M.D.Baker$^2$,
M.Ballintijn$^4$,
D.S.Barton$^2$,
R.R.Betts$^6$,
A.A.Bickley$^7$,
R.Bindel$^7$,
A.Budzanowski$^3$,
W.Busza$^4$,
A.Carroll$^2$,
Z.Chai$^2$,
V.Chetluru$^6$,
M.P.Decowski$^4$,
E.Garc\'{\i}a$^6$,
T.Gburek$^3$,
N.George$^2$,
K.Gulbrandsen$^4$,
S.Gushue$^2$,
C.Halliwell$^6$,
J.Hamblen$^8$,
G.A.Heintzelman$^2$,
C.Henderson$^4$,
I.Harnarine$^6$,
D.J.Hofman$^6$,
R.S.Hollis$^6$,
R.Ho\l y\'{n}ski$^3$,
B.Holzman$^2$,
A.Iordanova$^6$,
E.Johnson$^8$,
J.L.Kane$^4$,
N.Khan$^8$,
W.Kucewicz$^6$,
P.Kulinich$^4$,
C.M.Kuo$^5$,
W.Li$^4$,
W.T.Lin$^5$,
C.Loizides$^4$,
S.Manly$^8$,
A.C.Mignerey$^7$,
R.Nouicer$^{2,6}$,
A.Olszewski$^3$,
R.Pak$^2$,
I.C.Park$^8$,
C.Reed$^4$,
L.P.Remsberg$^2$,
M.Reuter$^6$,
E.Richardson$^7$,
C.Roland$^4$,
G.Roland$^4$,
L.Rosenberg$^4$,
J.Sagerer$^6$,
P.Sarin$^4$,
P.Sawicki$^3$,
I.Sedykh$^2$,
W.Skulski$^8$,
C.E.Smith$^6$,
M.A.Stankiewicz$^2$,
P.Steinberg$^2$,
G.S.F.Stephans$^4$,
A.Sukhanov$^2$,
A.Szostak$^2$,
J.-L.Tang$^5$,
M.B.Tonjes$^7$,
A.Trzupek$^3$,
C.Vale$^4$,
G.J.van~Nieuwenhuizen$^4$,
S.S.Vaurynovich$^4$,
R.Verdier$^4$,
G.I.Veres$^4$,
P.Walters$^8$,
E.Wenger$^4$,
D.Willhelm$^2$,
F.L.H.Wolfs$^8$,
B.Wosiek$^3$,
K.Wo\'{z}niak$^3$,
A.H.Wuosmaa$^1$,
S.Wyngaardt$^2$,
B.Wys\l ouch$^4$\\
\vspace{3mm}
\small
$^1$~Argonne National Laboratory, Argonne, IL 60439-4843, USA\\
$^2$~Brookhaven National Laboratory, Upton, NY 11973-5000, USA\\
$^3$~Institute of Nuclear Physics PAN, Krak\'{o}w, Poland\\
$^4$~Massachusetts Institute of Technology, Cambridge, MA 02139-4307, USA\\
$^5$~National Central University, Chung-Li, Taiwan\\
$^6$~University of Illinois at Chicago, Chicago, IL 60607-7059, USA\\
$^7$~University of Maryland, College Park, MD 20742, USA\\
$^8$~University of Rochester, Rochester, NY 14627, USA\\
}

\runtitle{}
\runauthor{}

\begin{document}

\maketitle

\begin{abstract}
The transverse momentum ($p_T$) spectra of identified charged particles have been measured at very low and intermediate transverse momenta in Au+Au collisions at $\sqrt{s_{NN}}$ = 62.4 GeV and d+Au collisions at $\sqrt{s_{NN}}$ = 200 GeV using the PHOBOS detector at RHIC. New results on charged particle production at very low $p_T$ in central Au+Au collisions at $\sqrt{s_{NN}}$ = 200 GeV in the centrality intervals 0-6\% and 6-15\% are presented. A comparison of the PHOBOS low-$p_T$ data with predictions of a recent optical model is shown. The shapes of $m_T$ spectra for d+Au and Au+Au collisions are compared.
\end{abstract}

\section{Introduction}
In nucleus-nucleus collisions, an enhanced production of low-$p_T$ particles could signal new long-wavelength physics phenomena~\cite{Busza,Bjorken}. It is also expected that yields of particles with higher masses, like protons and antiprotons can be modified due to collective transverse expansion of the system~\cite{Kolb,Schnedermann}. Measurements at very low $p_T$ can also provide a critical test for models predicting a pronounced modification of the low-$p_T$ particle emission pattern, e.g.~\cite{cramer}.

The PHOBOS experiment has the unique capability to measure charged particles at transverse momenta as low as 30, 90 and 140 MeV/c for charged pions, kaons and for protons and antiprotons, respectively, using a multi-layer, magnetic spectrometer. Yields at very low transverse momenta are determined using a reconstruction procedure developed to look for particles which range out in the fifth silicon layer of the PHOBOS spectrometer. A description of the ''stopping algorithm'' is presented in \cite{lowpt}. At intermediate $p_T$, particle momentum and charge are obtained from the curvature of particle trajectories in a 2T magnetic field and particle identification is provided by the specific energy loss ($dE/dx$) in the spectrometer and by Time-of-Flight detectors. Details on tracking, particle identification, event selection and centrality determination in the PHOBOS detector can be found in \cite{WP}.

\section{$\bf p_T$ spectra in Au+Au collisions at  $\bf \sqrt{s_{NN}}$ = 62.4 GeV}

\vspace{-0.7cm}
\begin{figure}[hb]
\begin{minipage}[t]{7cm}
\hspace{0.3cm} The preliminary particle yields for $\pi^{\pm}$, $K^{\pm}$, $p$ and $\bar{p}$ are presented in Fig.~\ref{spect624} for three centrality intervals: 0-15\%, 15-30\% and 30-50\%. The data are corrected for detector effects (acceptance, efficiency, momentum resolution) and background particles including feed-down from weak   decays and secondary particles produced in the beam pipe and detector material.  The rapidity coverage of measured yields extends from about 0.4 to 1.4 for  $\pi^{\pm}$, from 0.2 to 1.2 for $K^{\pm}$  and from 0.2 to 1.1 for $p$ and $\bar{p}$. The preliminary results on antiparticle to particle ratios have been obtained for the 15\% most central collisions. The results of 0.84 $\pm$ 0.02(stat.) $\pm$ 0.08(syst.) for $K^{-}/K^{+}$ and  0.37 $\pm$ 0.01(stat.) $\pm$ 0.06(syst.) for $\bar{p}/p$ fit smoothly into the energy evolution of antiparticle to particle ratios from the AGS up to the highest RHIC energy. 

\hspace{0.3cm} Low-$p_T$ yields of ($\pi^{+}+\pi^{-}$), ($K^{+}+K^{-}$) and  ($p+\bar{p}$) near mid-rapidity in Au+Au collisions at $\sqrt{s_{NN}}$ = 62.4 GeV, corrected for detector effects and background particles, are shown in Fig.~\ref{lowspect624} in the same centrality bins. One can see that ($K^{+}+K^{-}$) and  ($p+\bar{p}$) yields are quite consistent with
\end{minipage}
\hspace{\fill}
\begin{minipage}[t]{8.2cm}
\vspace{0.5cm}
\includegraphics[width=8.3cm]{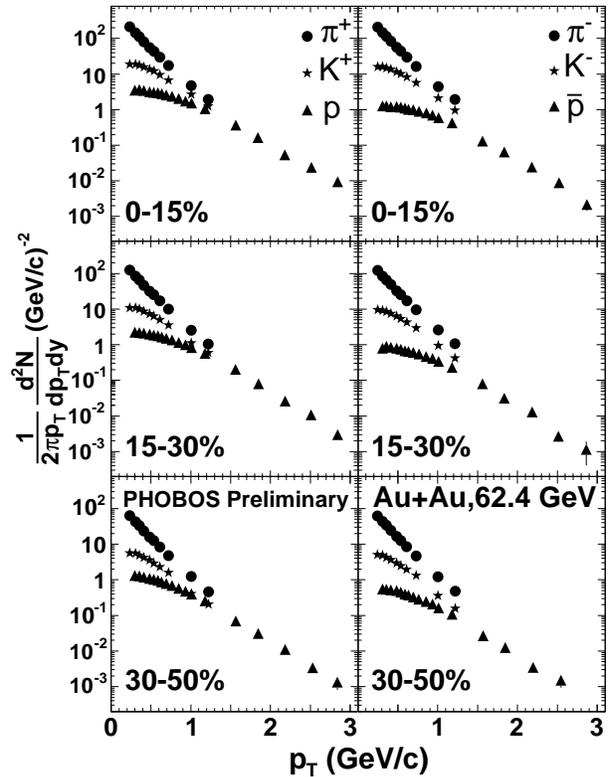}
\vspace{-1.5cm}
\caption{$p_T$ spectra of $\pi^{\pm}$, $K^{\pm}$, p and  $\bar{p}$ near mid-rapidity in Au+Au collisions at  $\sqrt{s_{NN}}$ = 62.4 GeV.}
\label{spect624}
\end{minipage}
\end{figure}

\vspace{-1cm}
\noindent 
 extrapolations of blast wave functions (BWF)~\cite{Schnedermann} fitted to the spectra at higher transverse momenta. Some disagreement between the measured yield of pions and BWF at low $p_T$ could be attributed to a contribution from resonances which is not included in  the model. A similar behavior was observed for $p_T$ yields  measured in the 15\% most central Au+Au collisions at $\sqrt{s_{NN}}$ = 200 GeV~\cite{lowpt}, indicating that at both energies no significant enhancement of particle production is observed at very low $p_T$. Also, a flattening of the ($p+\bar{p}$) spectra down to very low transverse momentum is observed. This could be a consequence of collective transverse expansion of the medium created in heavy ion collisions at RHIC.

\begin{figure}[ht]
\begin{minipage}[hb]{5.0cm}
\vspace{-.3cm}
\includegraphics[width=5cm]{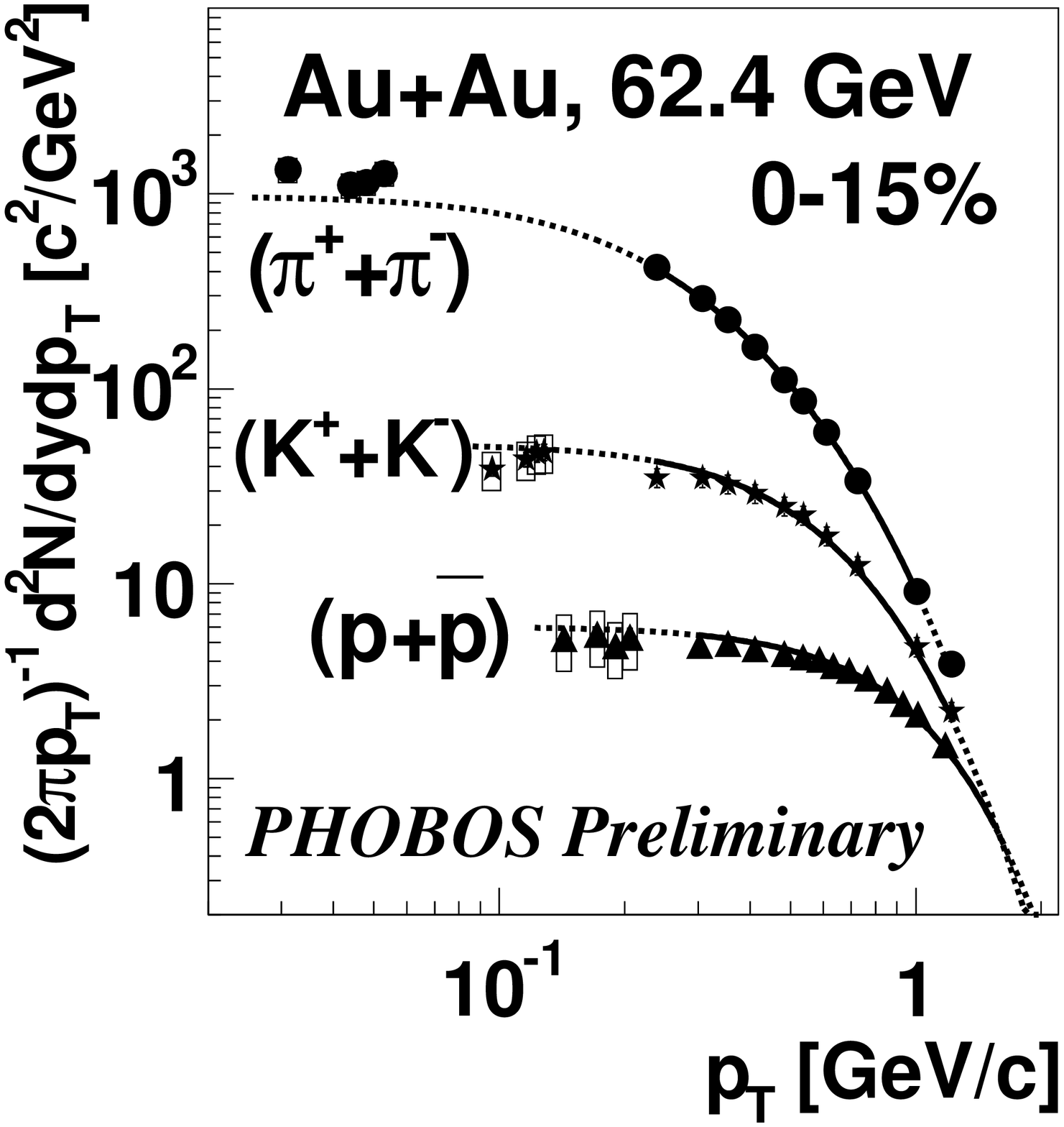}
\end{minipage}
\begin{minipage}[hb]{5.0cm}
\vspace{-.3cm}
\includegraphics[width=5cm]{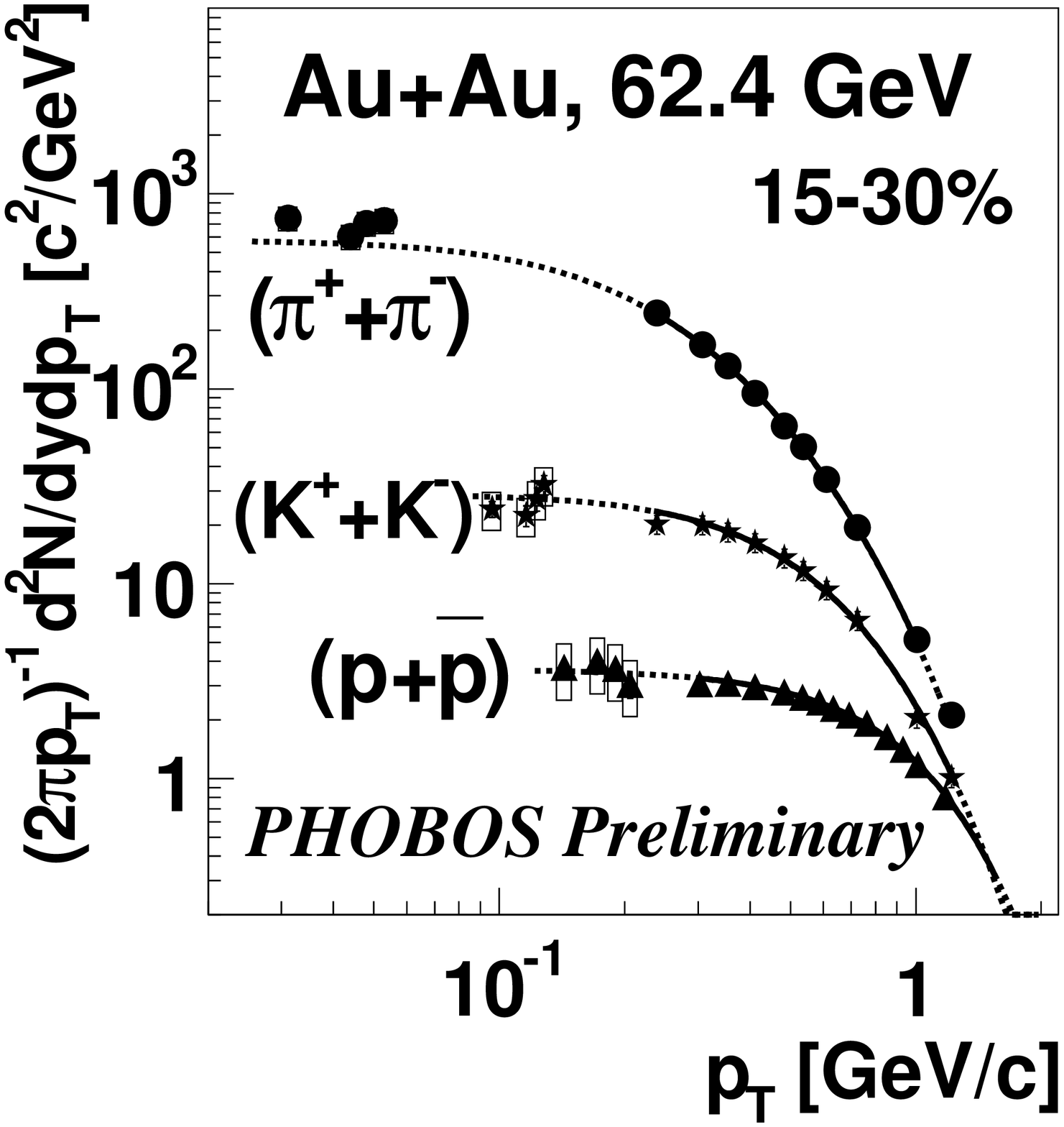}
\end{minipage}
\begin{minipage}[hb]{5.0cm}
\vspace{-.3cm}
\includegraphics[width=5cm]{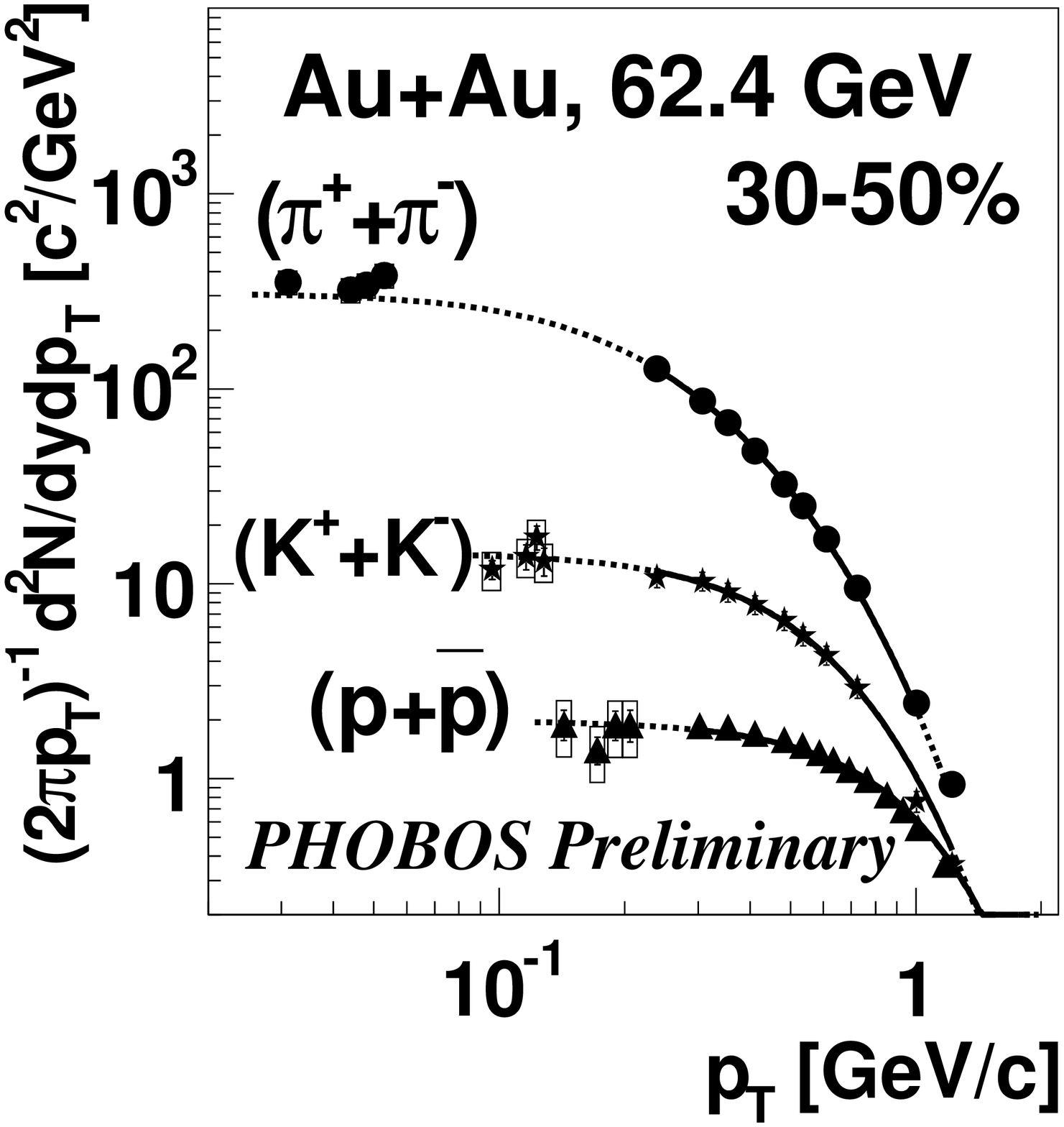}
\end{minipage}
\vspace{-0.5cm}
\caption{ ($\pi^{+}+\pi^{-}$), ($K^{+}+K^{-}$) and  ($p+\bar{p}$) yields at very low $p_T$ in Au+Au collisions at  $\sqrt{s_{NN}}$ = 62.4 GeV. Blast wave fits to the intermediate $p_T$ data (solid lines) are extrapolated to low $p_T$ (dashed lines).}
\label{lowspect624}
\end{figure}
 
\vspace{-1cm}
\section{Low-$\bf p_T$ yields in central Au+Au collisions at  $\bf \sqrt{s_{NN}}$ = 200 GeV}

\begin{figure}[b]
\begin{minipage}[t]{5cm}
\vspace{-.5cm}
\includegraphics[width=4.5cm]{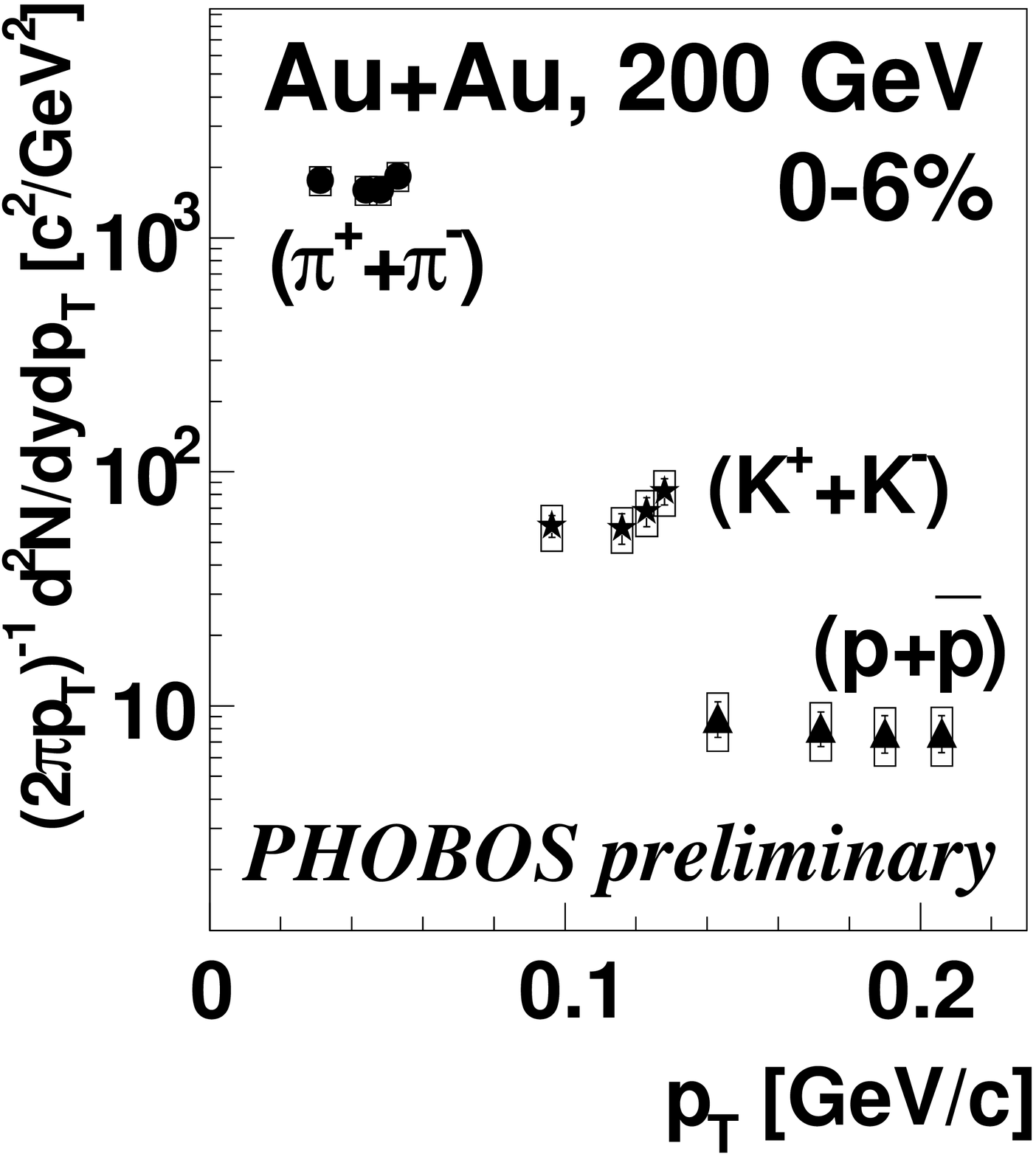}
\end{minipage}
\begin{minipage}[t]{5cm}
\vspace{-.5cm}
\includegraphics[width=4.5cm]{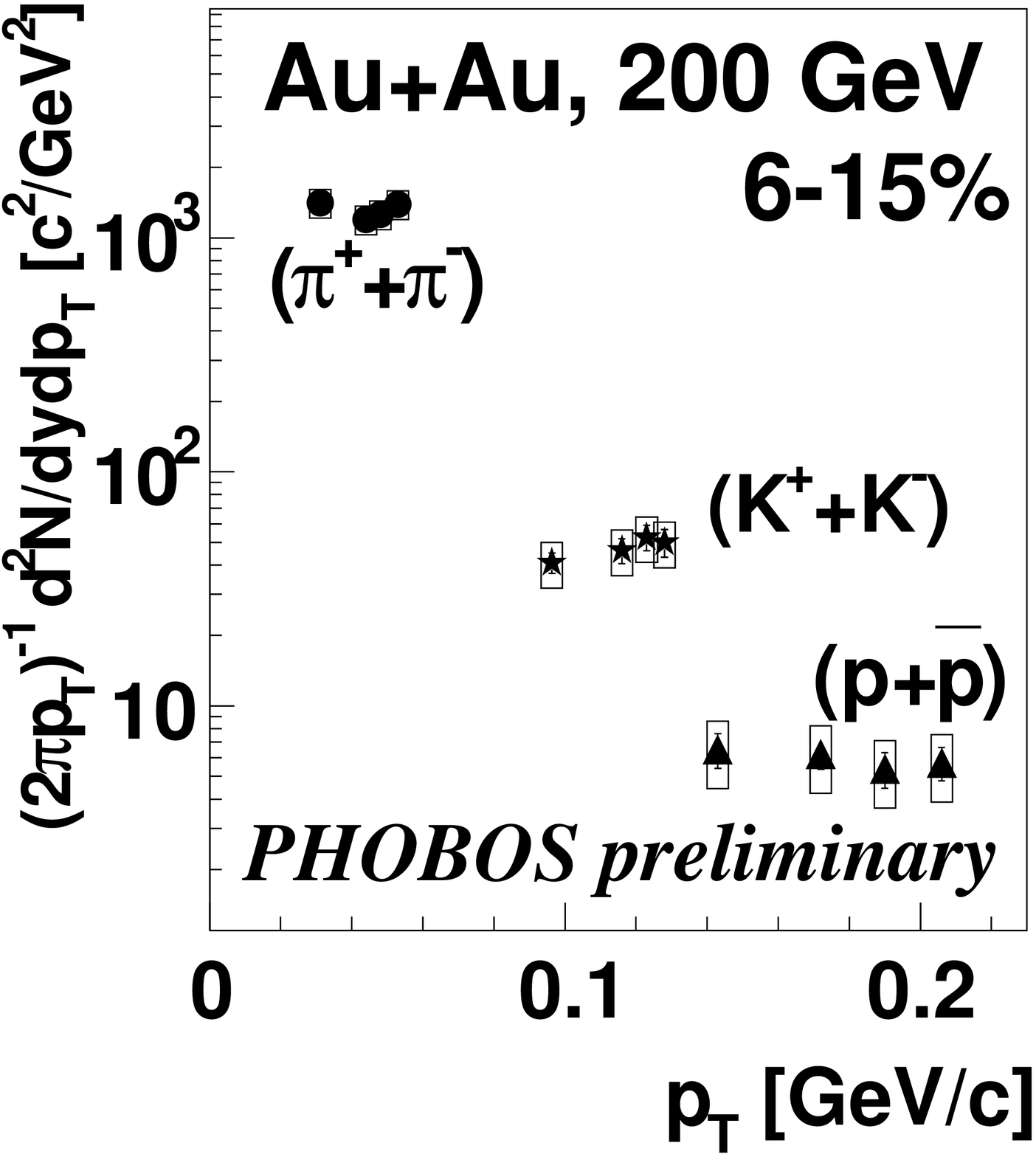}
\end{minipage}
\hspace{0.3cm}\begin{minipage}[t]{5cm}
\vspace{-.5cm}
\includegraphics[width=5.1cm]{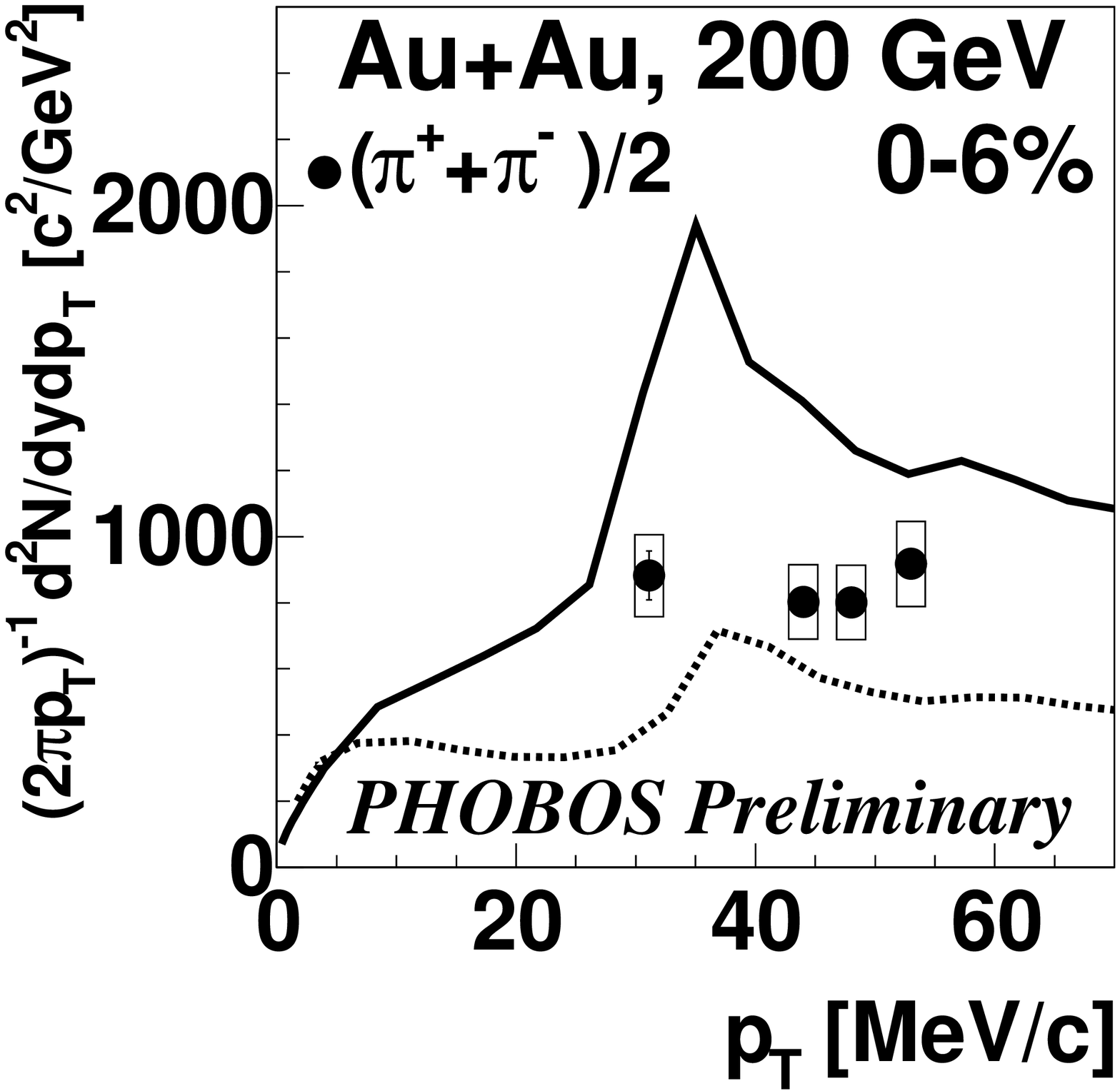}
\end{minipage}
\begin{minipage}[b]{10cm}
\vspace{-0.8cm}
\caption{($\pi^{+}+\pi^{-}$), ($K^{+}+K^{-}$) and ($p+\bar{p}$) yields at very low $p_T$ in 0-6\% and 6-15\% central Au+Au collisions at $\sqrt{s_{NN}}$ = 200 GeV. }
\label{spect200}
\end{minipage}
\hspace{0.8cm}\begin{minipage}[t]{5.2cm}
\vspace{-2.3cm}
\caption{Optical model predictions \cite{cramer,cramerE} for pion spectra at low $p_T$ compared to PHOBOS data (see text for details). }
\label{cramer}
\end{minipage}
\end{figure}

The spectra of ($\pi^{+}+\pi^{-}$), ($K^{+}+K^{-}$) and  ($p+\bar{p}$) at very low transverse momentum in the 15\% most central Au+Au collisions at $\sqrt{s_{NN}}$ = 200 GeV, measured in the PHOBOS experiment, are presented in \cite{lowpt}. In order to confront the extrapolations from a recent optical model ~\cite{cramer}, which were available only for more central collisions, with measurements, the published data sample was split into two finer centrality bins. Fig.~\ref{spect200} shows the $p_T$ yields, corrected for detector effects and background particles, measured in the centrality intervals 0-6\% and 6-15\%. In Fig.~\ref{cramer}, the pion yield measured in the 6\% most central Au+Au collisions is compared to the optical model predictions for the spectrum of negative pions at mid-rapidity. The originally published extrapolation is shown by the dashed curve, while the solid curve depicts the recently modified model calculations \cite{cramer,cramerE}.

\begin{figure}[t]
\hspace{1.5cm}
\begin{minipage}[t]{5.5cm}
\includegraphics[width=5.5cm]{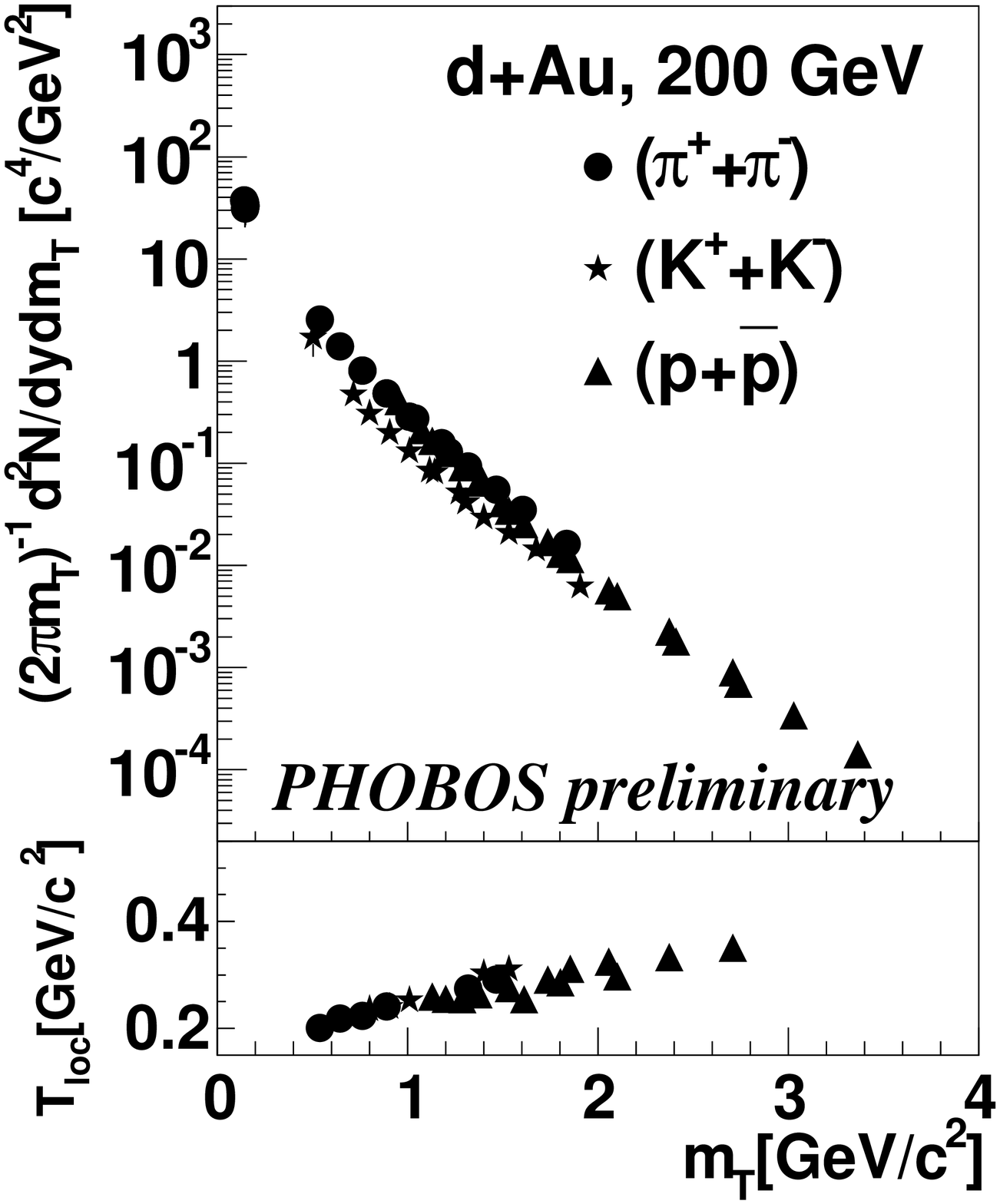}
\end{minipage}
\hspace{1cm}
\begin{minipage}[t]{5.5cm}
\includegraphics[width=5.5cm]{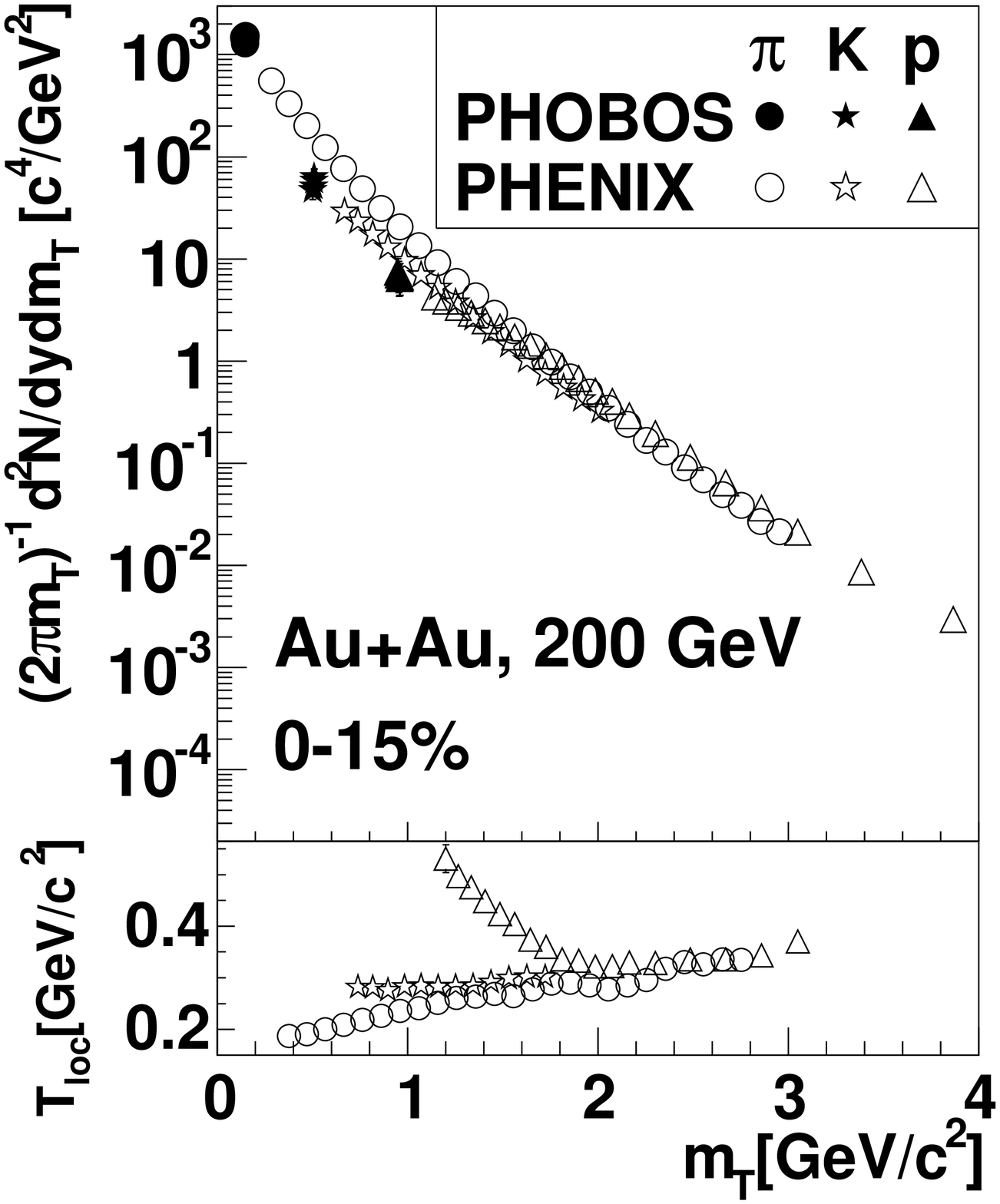}
\end{minipage}\\
\center\begin{minipage}[t]{14.5cm}
\vspace{-2cm}
\caption{$m_T$ spectra of ($\pi^{+}+\pi^{-}$), ($K^{+}+K^{-}$) and  ($p+\bar{p}$) at very low and intermediate $p_T$ measured in d+Au (left plot) and central Au+Au collisions (right plot). Inverse local slope parameters of $m_T$ spectra are shown in the lower panels of each figure.}
\label{dAu}
\end{minipage}
\end{figure}

\section{$\bf m_T$ scaling}

It is interesting to compare the particle yields at very low and intermediate $p_T$ in d+Au and central Au+Au collision at the same energy of $\sqrt{s_{NN}}$ = 200 GeV. Yields of ($\pi^{+}+\pi^{-}$), ($K^{+}+K^{-}$) and ($p+\bar{p}$) in d+Au collisions, corrected for detector effects and background particles, are shown in Fig.~\ref{dAu}. One can see that in d+Au collisions  ($\pi^{+}+\pi^{-}$) and ($p+\bar{p}$) $m_T$ spectra are similar while the ($K^{+}+K^{-}$) spectrum is systematically lower (by a factor of about 2) due to strangeness suppression. The $m_T$ spectra for the 15\% most central Au+Au collisions measured by the PHOBOS \cite{lowpt} and PHENIX \cite{Phenix} experiments at very low and intermediate tranverse momenta, respectively, are also shown in Fig.~\ref{dAu}. In order to compare the shapes of the $m_T$ spectra, inverse local slope parameters were calculated by fitting locally exponential functions to each spectrum (see bottom panels of Fig.~\ref{dAu}). We can see that for d+Au collisions local slopes are similar for all particle species both at low and intermediate $p_T$. In contrast, Au+Au spectral shapes are similar at higher transverse masses ($m_T>$~1.7 GeV) while at low $m_T$ a flattening of ($K^{+}+K^{-}$) and ($p+\bar{p}$) spectra is observed. This flattening of the ($p+\bar{p}$) spectrum is significantly stronger than the one observed for the spectra of charged kaons. One can also see that the $m_T$ dependence of the local slopes of the ($\pi^{+}+\pi^{-}$) $m_T$ spectrum for Au+Au collision is consistent with that found for the local slopes of  ($\pi^{+}+\pi^{-}$),   ($K^{+}+K^{-}$) and ($p+\bar{p}$) spectra in d+Au collisions.

\section{Summary}
\vspace{.1cm}
Yields of pions, kaons and protons and antiprotons at very low and intermediate $p_T$ near mid-rapidity in Au+Au collisions at $\sqrt{s_{NN}}$ = 62.4 and 200 GeV, measured in the PHOBOS experiment,  indicate that there is no evidence for enhanced production of particles at very low transverse momentum. The pion low-$p_T$ data can constrain the recent optical model predictions. A significant flattening of the ($p+\bar{p}$) $m_T$ spectrum down to very low $p_T$ is observed in central Au+Au collisions at both energies which could be a consequence of the collective transverse expansion of the system. In d+Au collisions, no flattening is observed and the shapes of the $m_T$ spectra are similar at very low and intermediate $p_T$. \\[1mm]

%
%
%
%
This work was partially supported by U.S. DOE grants 
DE-AC02-98CH10886,
DE-FG02-93ER40802, 
DE-FC02-94ER40818,  
DE-FG02-94ER40865, 
DE-FG02-99ER41099, and
W-31-109-ENG-38, by U.S. 
NSF grants 9603486, 
0072204,            
and 0245011,        
by Polish KBN grant 1-P03B-062-27(2004-2007),
by NSC of Taiwan Contract NSC 89-2112-M-008-024, and
by Hungarian OTKA grant (F 049823).

\end{document}